\documentclass[english,draft,letterpaper]{smfart}
\usepackage{latexsym,amssymb}%

\usepackage{graphicx}  

\usepackage{color}

\usepackage{a4}
\usepackage{amssymb,latexsym}
\usepackage[mathscr]{eucal}
\usepackage{cite}
\usepackage{url}

\DeclareFontFamily{OT1}{rsfs}{} \DeclareFontShape{OT1}{rsfs}{m}{n}{ <-7> rsfs5 <7-10> rsfs7 <10-> rsfs10}{}
\DeclareMathAlphabet{\mycal}{OT1}{rsfs}{m}{n}
\def\scri{{\mycal I}}

\newcommand{\mg}{\mathring g}%
\newcommand{\mk}{\mathring k}%
\newcommand{\mnju}{\mathring \nu}%
\newcommand{\mmr}{\mathring m_r}%

\newcommand{\kk}[1]{}

\newcommand{\zD}{\mathring D}

\newcommand{\Span}{\mathrm{Span}}

\newcommand{\bmcM}{\,\,\,\,\widetilde{\!\!\!\!\mcM}}

\newcommand{\eean}{\nonumber\end{eqnarray}}

\newcommand{\mcMext}{\Mext}

\def\KK{\phi^K}
\def\K0{\phi^{K_0}}

\def\X.{\phi^{X}\cdot}

\newcommand{\hmcM}{\,\,\,\widehat{\!\!\!\mcM}}

{\catcode `\@=11 \global\let\AddToReset=\@addtoreset}
\AddToReset{equation}{section}
\renewcommand{\theequation}{\thesection.\arabic{equation}}

\newcommand{\fourg}{{\mathfrak g }}

\newcommand{\nopcite}[1]{}

\newcommand{\del}{\partial}

  {\renewcommand{\theenumi}{\roman{enumi}}
   \renewcommand{\labelenumi}{(\theenumi)}}

\newcommand{\const}{\mathrm{const}}

\newcommand{\mcD}{{\mycal D}}

\newcommand{\nablash}{\nabla{\kern -.75 em
     \raise 1.5 true pt\hbox{{\bf/}}}\kern +.1 em}
\newcommand{\Deltash}{\Delta{\kern -.69 em
     \raise .2 true pt\hbox{{\bf/}}}\kern +.1 em}
\newcommand{\Rslash}{R{\kern -.60 em
     \raise 1.5 true pt\hbox{{\bf/}}}\kern +.1 em}

\newcommand{\mcO}{{\mycal O}}
\newcommand{\mcT}{{\mycal T}}
\newcommand{\mcU}{{\mycal U}}

\newcommand{\hyp}{{\mycal S}}

\newcommand{\threeg}{\gamma}

\newcommand{\mcM}{{\mycal M}}

\newcommand{\bea}{\begin{eqnarray}}
\newcommand{\beaa}{\begin{eqnarray*}}
\newcommand{\bean}{\begin{eqnarray}\nonumber}

\newcommand{\bel}[1]{\begin{equation}\label{#1}}
\newcommand{\beal}[1]{\begin{eqnarray}\label{#1}}
\newcommand{\beadl}[1]{\begin{deqarr}\label{#1}}
\newcommand{\eeadl}[1]{\arrlabel{#1}\end{deqarr}}
\newcommand{\eeal}[1]{\label{#1}\end{eqnarray}}
\newcommand{\eead}[1]{\end{deqarr}}
\newcommand{\eea}{\end{eqnarray}}
\newcommand{\eeaa}{\end{eqnarray*}}

\newcommand{\be}{\begin{equation}}
\newcommand{\ee}{\end{equation}}

\newcommand{\tr}{\mbox{\rm tr}\,}

\newcommand{\eq}[1]{\eqref{#1}}
\newcommand{\Eq}[1]{Equation~(\ref{#1})}

\newtheorem{defi}{\sc Coco\rm}[section]

\newtheorem{Theorem}[defi]{\sc Theorem\rm}

\newtheorem{Proposition}[defi]{\sc Proposition\rm}

\newtheorem{Example}[defi]{{\sc Example}\rm}
\newtheorem{Remark}[defi]{{\sc Remark}\rm}
\newtheorem{Remarks}[defi]{{\sc Remarks}\rm}

\def \R {\Reel}

\newcommand{\mcL}{{\mycal L}}

\newcounter{mnotecount}[section]

\renewcommand{\themnotecount}{\thesection.\arabic{mnotecount}}

\newcommand{\mnote}[1]
{\protect{\stepcounter{mnotecount}}$^{\mbox{\footnotesize $%
\!\!\!\!\!\!\,\bullet$\themnotecount}}$ \marginpar{
\raggedright\tiny\em $\!\!\!\!\!\!\,\bullet$\themnotecount: #1} }

\newcommand{\ednote}[1]{}

\definecolor{bluem}{rgb}{0,0,0.5}

\definecolor{mycolor}{cmyk}{0.5,0.1,0.5,0}
\definecolor{michel}{rgb}{0.5,0.9,0.9}

\definecolor{turquoise}{rgb}{0.25,0.8,0.7}
\definecolor{bluem}{rgb}{0,0,0.5}

\definecolor{MDB}{rgb}{0,0.08,0.45}
\definecolor{MyDarkBlue}{rgb}{0,0.08,0.45}

\definecolor{MLM}{cmyk}{0.1,0.8,0,0.1}
\definecolor{MyLightMagenta}{cmyk}{0.1,0.8,0,0.1}

\definecolor{HP}{rgb}{1,0.09,0.58}

\newcommand{\Sext}{\hyp_{\mathrm{ext}}}
\newcommand{\Mext}{\mcM_{\mathrm{ext}}}

\newcommand{\doc}{\langle\langle \mcMext\rangle\rangle}
\newcommand{\docl}{\langle\langle\mcMext^\lambda\rangle\rangle}
\newcommand{\docls}{\langle\langle \mcMext^{\lambda_*}\rangle\rangle}

\def\emph#1{{\it #1}}
\def\textbf#1{{\bf #1}}

\def\ga{\gamma}

\def\tr{\mbox{tr}}

\def\AA{{\mycal  A}}

\def\KK{{\mycal  K}}

\def\L{{\bf L}}
\def\O{{\bf O}}

\def\R{{\mathbb R}}

\def\K{{\bf K}}

\def\T{{\Bbb T}}

\def\2{{\overline 2}}

\newcommand{\beqa}{\begin{eqnarray}}
\newcommand{\eeqa}{\end{eqnarray}}

\begin {document}

\frontmatter
\author [P.T.~Chru\'sciel]{Piotr T.~Chru\'sciel}
\address {Mathematical Institute and Hertford College, Oxford; LMPT,
F\'ed\'eration Denis Poisson, Tours}
 \email{chrusciel@maths.ox.ac.uk}
\urladdr {www.phys.univ-tours.fr/$\sim$piotr}

\author [G.J.~Galloway]{Gregory J.~Galloway\thanks{Supported in part by  NSF grant DMS-0708048}}
\address {University of Miami, Coral Gables}
 \email {galloway@math.miami.edu}
\urladdr {www.math.miami.edu/~galloway}
\author [D.~Solis]{Didier Solis}
\address {Universidad Autonoma de Yucatan, Merida}
 \email {didier.solis@uady.mx}

\title[Topological censorship for Kaluza-Klein space-times]{Topological censorship for Kaluza-Klein space-times}

\begin {abstract}
The standard topological censorship theorems require asymptotic
hypotheses which are too restrictive for several situations of
interest. In this paper we prove a version of topological
censorship under significantly weaker conditions, compatible
e.g. with solutions with Kaluza-Klein asymptotic behavior. In
particular we prove simple connectedness of the quotient of the
domain of outer communications by the group of symmetries for
models which are asymptotically flat, or asymptotically anti-de
Sitter, in a Kaluza-Klein sense. This allows one, e.g., to
define the twist potentials needed for the reduction of the
field equations in uniqueness theorems. Finally, the methods
used to prove the above are used to show that weakly trapped
compact surfaces cannot be seen from Scri.
\end {abstract}
\maketitle

\setcounter{tocdepth}{3}
\tableofcontents

\mainmatter

\section{Introduction}
\label{Sintro}

A  restriction on the topology of domains of outer
communications is provided by the \emph{topological censorship
principle}~\cite{FriedmanSchleichWitt}, which says that causal
curves originating from, and ending in a simply connected
asymptotic region do not see any non-trivial topology, in the
sense that they can be deformed to a curve entirely contained
within the asymptotic region. The result is one of the key
steps in the black holes uniqueness theorems (see, e.g.,
\cite{ChCo} and references therein). Precise statements to this
effect have been established in the literature under various
conditions~\cite{GSWW2,GSWW,ChWald,galloway-topology,Galloway:fitopology,Jacobson:venkatarami}.
Of particular relevance to our  work
is~\cite{Galloway:fitopology}, where topological censorship is
reduced to a \emph{null convexity} condition of timelike
boundaries. The first main result of this work is the proof
that the conditions
of~\cite{Galloway:fitopology} can be replaced by the
considerably weaker hypothesis, that the timelike boundaries are
inner future and past trapped, as defined below.

The need for this generalisation arises when studying the
topology of higher dimensional domains of outer communications
invariant under isometry groups. Recall that for asymptotically
flat stationary space-times, whatever the space-dimension $n\ge
3$, simple connectedness holds for globally hyperbolic domains
of outer communications satisfying the null energy condition.
Indeed, the analysis
in~\cite{ChWald,galloway-topology,Galloway:fitopology,FriedmanSchleichWitt},
carried-out there in dimension $3+1$, is independent of
dimensions. However, there exist significant higher dimensional
solutions which are asymptotically flat \emph{in a Kaluza-Klein
sense} and which are \emph{not} simply connected in general, as
demonstrated by Schwarzschild$\times \T^m$ ``black branes".

Now, whenever simple connectedness fails, the twist potentials
 characterising the Killing vectors might fail to exist,  and
the whole reduction process~\cite{CarterlesHouches,BMG}, that
relies on the existence of those potentials, breaks down. Our
next main result  is the proof that the quotient space
$\doc/G_s$ remains simply connected for $KK$--asymptotically
flat, or $KK$--asymptotically adS models, which is sufficient
for existence of twist potentials under mild conditions on
$\doc/G_s$,  and has some further significant applications in
the study of the problem at hand, see~\cite{ChHighDim}. Here a
uniformity-in-time condition is assumed on the asymptotic decay
of the metric, which will certainly be satisfied by stationary
solutions.

 It turns out that the methods here are well suited to address
the following: it is a well established fact in general
relativity that compact future trapped surfaces cannot be seen
from infinity. The weakly trapped  counterpart of this has
often been used in the literature,
without a satisfactory justification available.%
\footnote{This fact has been first brought to our attention by David Maxwell.}
Our last main result here is the proof that borderline
invisibility does indeed hold under appropriate global hypotheses.

\section{Preliminaries}
 \label{Sprelim}

All manifolds are
assumed to be Hausdorff and paracompact.
We use the signature $(-,+,\ldots,+)$, and all space-times have
dimension greater than or equal to three.

\subsection{Trapped surfaces}
 \label{sSts}
\renewcommand{\fourg}{g}%
 Let
$(\mcM,\fourg)$ be a space-time, and consider a spacelike manifold
$S\subset \mcM$ of co-dimension two. Assume that there exists a
smooth unit spacelike vector field $n$ normal to $S$.  If $S$
is a two-sided boundary of a set contained within a spacelike
hypersurface $\hyp$, we shall always choose $n$ to be the
outwards directed normal tangent to $\hyp$; this justifies the
name of \emph{outwards normal} for $n $. If $S\subset \{r=R\}$
in a $KK$--asymptotically flat or adS space-time, as defined in
Section~\ref{sKKa} below, then the outwards normal is defined to
be the one for which $n(r)>0$.

At every point $p\in S$ there exists then a unique future
directed null vector field $n^+$ normal to $S$ such that
$\fourg(n,n^+)=1$, which we shall call the \emph{outwards
future null normal} to $S$. The \emph{inwards future null
normal} $n^-$ is defined   by the requirement that $n^-$ is
null, future directed, with $\fourg(n,n^-)=-1$.

We define the \emph{null future inwards and outwards mean curvatures $\theta^\pm$ of $S$}   as
\bel{thetdef}
 \theta^\pm :=\tr_\gamma( \nabla n^\pm)
 \;,
\ee
where $\gamma$ is the metric induced on $S$. In \eq{thetdef}
the symbol $n^\pm$ should be understood as representing any
extension of the null normals $n^\pm$ to a neighborhood of $S$,
and the definition is independent of the extension chosen.

We shall say that $S$ is \emph{weakly outer future trapped} if
 $\theta^+\le 0$.  The notion of
  \emph{weakly inner future trapped} is defined by requiring
 $\theta^-\le 0$. A similar notion of \emph{weakly outer or
 inner past trapped} is defined by changing $\le$ to $\ge$ in
 the defining inequalities above. We will say \emph{outer
 future trapped} if $\theta^+< 0$, etc.
 One can also think of such conditions as \emph{mean null convexity conditions}.

Let $\mcT$ be a smooth timelike hypersurface in $\mcM$ with   a
globally defined smooth field $n$ of   unit normals to $\mcT$.
We shall say that $\mcT$ is \emph{weakly outer past trapped}
with respect to a time function $t$   if the level sets of $t$
on $\mcT$ are  {weakly outer past trapped}. A similar
definition is used for the notion of \emph{weakly outer future
trapped} timelike hypersurfaces, etc.

\subsection{Spacetimes with timelike boundary}

A space-time $(\mcM,\fourg)$ with timelike boundary $\mcT$ will
be said \emph{globally hyperbolic} if $(\mcM,\fourg)$ is
strongly causal and if for all $p,q\in \mcM$ the sets
$J^+(p)\cap J^-(q)$ are empty or compact. In this case a
hypersurface $\hyp$ is said to be a \emph{Cauchy surface} if
$\hyp$ is met by every inextendible causal curve precisely
once. A smooth
function $t$ is said to be a \emph{Cauchy time
function} if it ranges over $\R$, if $\nabla t$ is timelike
past directed,   and if all level sets are Cauchy surfaces.

As an example, let $\mcT$ be any sufficiently distant level set
of the usual radial coordinate $r$ in   Schwarzschild
space-time. Then $\mcT$ is both   inner   future and past
trapped, see \eq{trappingT} below.

The causal theory of spacetimes with timelike boundary has been
studied in detail in~\cite{Solis}. Many important results are
shown to be valid in this context. For instance, chronological
future and past sets are open and global hyperbolicity as
defined above implies causal simplicity. The following basic
property of Cauchy surfaces holds as well, and is stated for
future reference.

\begin{Proposition}
\label{PCSS}If $\hyp$ is a Cauchy surface and $K$ is a compact subset of $\mcM$ then $J^+(K) \cap J^-(\hyp)$ and $J^+(K) \cap
 \hyp $ are compact.
 \qed
\end{Proposition}


\section{Topological censorship for spacetimes with timelike boundary}
 \label{stcghsc}

We have the following generalisation of~\cite[Theorem~1]{Galloway:fitopology}:

\begin{Theorem}\label{TGconv}
Let $t$ be a Cauchy time function  on a space-time
 $(\mcM,\fourg)$ with  timelike boundary $\mcT=\cup_{\alpha\in
 \Omega} \mcT_\alpha$, and  satisfying the null energy condition
 (NEC):
 \bel{NEC}
 R_{\mu\nu}X^\mu X^\nu \ge 0 \quad \mbox{\rm for all null vectors $X^\mu$}.
 \ee
 Suppose that there exists a component $\mcT_1$ of $\mcT$ with compact level sets $t|_{\mcT_1}$ such that
 $$ \mcT_1  \ \mbox{\rm{is weakly inner future trapped}}$$
  with respect to $t$. If
 $$\mbox{\rm{all connected components} $\mcT_\alpha$, $\alpha\ne 1$, \rm{of} $\mcT$ \rm{are} \rm{ inner past trapped}}$$
 with respect to $t$, then
$$
 J^+(\mcT_1)\cap J^-(\mcT_\alpha) =\emptyset \ \mbox{ for } \ \mcT_\alpha \ne  \mcT_1
 \;.
$$
\end{Theorem}

\begin{Remarks}{\rm 1. Nothing is assumed about the nature of the index set $\Omega$.

2.  The condition that at least one of the defining inequalities is strict is necessary. Indeed, let $\mcM'=\R\times \underbrace{S^1\times \ldots
\times S^1}_{\text{$n$ factors}}$ with a flat product metric, let $t$ be a standard coordinate on the $\R$ factor, and let $\varphi\in [0,2\pi]$ be a
standard angular coordinate on the first $S^1$ factor, then $\mcM=\{0\le \varphi \le \pi\}\subset \mcM'$ satisfies the hypotheses above except for the
strictness condition, and does not satisfy the conclusion. }
\end{Remarks}

Let $t$ be a time function on $\mcM$, and let $\gamma:[a,b]\to \mcM$ be a future directed causal curve. The \emph{time of flight $t_\gamma$ of $\gamma$} is defined as

$$
 t_\gamma= t(\gamma(b))-t(\gamma(a))
 \;.
$$
As a step in the proof of Theorem~\ref{TGconv}, we note:

\begin{Proposition}
\label{Pnotrapped0} Let $(\mcM,\fourg)$ be a globally
hyperbolic space-time satisfying the null energy condition
containing a past inwards trapped hypersurface $\mcT$. Let
$S\subset \mcM$ be future inwards weakly trapped.  Then there
are no future directed causal curves,  starting inwardly at
$S$, meeting $\mcT$ inwardly, and  minimising,  amongst nearby
causal curves, the time of flight between $S$ and $\mcT$.
\end{Proposition}

\begin{proof}
Suppose that the result is wrong, thus there exists a future
directed causal curve $\gamma:[a,b]\to \mcM$ with $\gamma(a)\in
S$, $\gamma(b)\in \mcT$, locally minimising the time of flight.
Standard considerations show that $\gamma$ is  a  null geodesic
emanating orthogonally from $S$ without $S$-conjugate points on
$[a,b)$. In particular $\dot J^+(S)$ is a smooth null
hypersurface near $\gamma{([a,b))}$. Let $t_+=t(\gamma(b))$,
set $S_+=\{t=t_+\}\cap \mcT$, then $\dot J^-(S_+)$ is a smooth null
hypersurface near $S_+$, that contains  a segment
$\gamma([b-\varepsilon,b])$. Moreover, $\dot J^-(S_+)$  lies to
the causal past of $\dot J^+(S)$ close to $\gamma$ since
otherwise we could construct a timelike curve from $S$ to
$S_+$ close to $\gamma$, thus violating the minimisation
character of $\gamma$. Since $\dot \gamma(a)$ is inwards
pointing and $\dot \gamma (b)$ outwards pointing, the
Raychaudhuri equation shows that the null divergence of $\dot
J^+(S)$ along $\gamma$ is non-positive whereas the null
divergence of $\dot J^-(S_+)$ is positive near $S_+$. For
points $\gamma(s)$, with $s\ne b$ but close to $b$, this
contradicts the maximum principle for null hypersurfaces
\cite{Galloway:splitting, GregCargese}, establishing the
result.
\end{proof}

\begin{Example}
 {\rm
 An example to keep in mind is the following: let $p,q\in
\R^{n+1}$ be two spatially separated points in Minkowski space-time $\R^{1,n}$. Let $S=\dot J^-(p)\cap \dot J^-{(q)  }$. The null generators of $\dot
J^-(p)$ and $\dot J^-(q)$ are converging, when followed to the future, and they meet $S$ normally, which shows that $S$ is an outwards and inwards
future trapped (non-compact) submanifold of $\R^{1,n}$. Of course, in this case the choice of ``inwards" and ``outwards" is a pure matter of
convention.

Let $n=3$, choose $p=(2,2,0,0)$, $q=(2,-2,0,0)$, let $\mcT$ be
the timelike surface $\mcT=\{r=1\}$, which is both future and
past  inwards trapped. The null achronal geodesic segments
$\gamma_\pm(t)=(t,\pm t,0,0)$, $ 0\le t<2$, are in $\dot
J^+(S)$ and, by symmetry considerations,  \emph{maximise} the
time of flight between $S$ and $\mcT$. Since $\dot J^+(S)$ lies
below $\dot J^-(\{t=0\}\cap \mcT)$, the argument in the proof
below, when applied to $\gamma_\pm$, does not lead to a
contradiction. But the example shows that the existence of an
achronal null geodesic segment between $S$ and $\mcT$ is
compatible with the hypotheses above. In particular ``locally
minimising" cannot be replaced by ``extremising". }
\end{Example}

{\noindent\sc Proof of Theorem~\ref{TGconv}:} Let $S$ be a weakly inner trapped
compact Cauchy surface of $\mcT_1$.  Suppose there exists a
causal curve $c$ from $S$ to a point $p$ in a different
component $\mcT_2$. Let $\hyp$ be the Cauchy surface $\{t =
t(p)\}$ for $M$. We want to construct a fastest null geodesic
from $S$ to $\mcT \setminus \mcT_1$; for this we need to show
that only finitely many components, ${\mcT_a}$, of $\mcT
\setminus \mcT_1$ meet the set $ A = J^+(S) \cap \hyp$, which
is compact by Proposition~\ref{PCSS}. Suppose to the contrary,
there are infinitely many of these components that meet $A$.
Then we obtain an infinite sequence of points $\{x_n\}$ in $A$,
each point in a different component. Since $A$ is compact we
can pass to a convergent subsequence, still called $\{x_n\}$,
such that $x_n \to x \in A$. Since $\mcT \cap \hyp$ is closed,
$x$ is in $\mcT$. But this
 contradicts the half-neighborhood property of
manifolds with boundary.

The time function $t$ on $M$ restricts to a time function on
$\mcT$. By the observation in the preceeding paragraph, the set
$  (\mcT\setminus \mcT_1)\cap J^+(S) \cap J^-(\hyp)$
  is compact
and thus we can now minimize $t$ on causal curves from $S$ to
$\mcT\setminus\mcT_1$ contained in the aforementioned compact
set to obtain a fastest causal curve $\gamma$ from $S$ to
$\cup_{a\ne 1} \mcT_a$. Since $t$ has been minimized, $\gamma$
meets $\mcT$ only at its endpoints, and hence must be a null
geodesic. This contradicts Proposition~\ref{Pnotrapped0}, and
establishes the result. See~\cite{Solis} for a {more detailed exposition}.
\qed

\medskip

We now proceed to establish a general topological censorship
result for globally hyperbolic spacetimes with timelike
boundary.

\begin{Theorem}\label{topcenwb}
Let $(\mcM,\fourg)$ be a spacetime with a connected timelike
boundary $\mcT$. Let
$$\langle\langle \mcT\rangle\rangle :=
I^+(\mcT)\cap I^-(\mcT)
$$
be the domain of  communications of $\mcT$. Further assume
$\langle\langle \mcT \rangle\rangle$ has a Cauchy time function
such that the level sets $t|_{\mcT}$ are compact.  If the NEC
holds on $\langle\langle\mcT\rangle\rangle$, and if $\mcT$ is
inner past trapped and weakly inner future trapped with respect
to $t$, then topological censorship holds, i.e., any causal
curve included within $\langle\langle\mcT\rangle\rangle$ with
end points on $\mcT$ can be deformed, keeping end points fixed,
to a curve included in $\mcT$.
\end{Theorem}

\begin{proof}
First notice that the inclusion $j\colon\mcT\hookrightarrow
\langle\langle\mcT\rangle\rangle$ induces a homomorphism of
fundamental groups $j_*\colon \pi_1(\mcT )\to
\pi_1(\langle\langle\mcT\rangle\rangle )$. Thus there exists a
covering $\pi\colon M\to \langle\langle\mcT\rangle\rangle$
associated to the subgroup  $j_*(\pi_1(\mcT ))$ of
$\pi_1(\langle\langle\mcT\rangle\rangle )$. This covering is
characterized as the largest covering of
$\langle\langle\mcT\rangle\rangle$ containing a homeomorphic
copy $\mcT_0$ of $\mcT$, that is, $\pi\vert_{\mcT_0}$ is a
homeomorphism onto $\mcT$~\cite{HHM}. Furthermore, this
covering has the property that the map $i_*\colon \pi_1(\mcT_0
)\to\pi_1(M )$ induced by the inclusion $i\colon\mcT_0\to M$ is
surjective. Endowing $M$ with the pullback metric $\pi^*(\fourg
)$ we get a globally hyperbolic spacetime with timelike
boundary $\pi^{-1}(\mcT )$. Now, let $\gamma\colon [a,b]\to
\langle\langle\mcT\rangle\rangle$ be a causal curve with
endpoints in $\mcT$. Lift $\gamma$ to $\gamma_0\colon [a,b]\to
M$ with $\gamma_0(a)\in \mcT_0$. By Theorem~\ref{TGconv} we
know that $\mcT_0$ can not communicate with any other component
of $\pi^{-1}(\mcT )$, hence $\gamma_0(b)\in \mcT_0$. As a
consequence, $\gamma_0$ is homotopic to a curve in $\mcT_0$ and
the result follows.
\end{proof}

As noted in~\cite{GSWW}, topological censorship can be viewed
as the statement that \emph{any} curve in
$\langle\langle\mcT\rangle\rangle$ with endpoints in $\mcT$ is
homotopic to a curve in $\mcT$, or equivalently that the map
$j_*\colon \pi_1(\mcT )\to
\pi_1(\langle\langle\mcT\rangle\rangle )$ is surjective. We
reproduce here the argument for completeness.

\begin{Theorem}\label{stopcen}
With the same hypotheses as above, the map $j_*\colon \pi_1(\mcT )\to \pi_1(\langle\langle\mcT\rangle\rangle )$ induced by the inclusion $j\colon \mcT\to \langle\langle\mcT\rangle\rangle$ is surjective.
\end{Theorem}

\begin{proof} Let $\pi\colon M \to {\langle\langle\mcT\rangle\rangle}$ be
the universal cover of ${\langle\langle\mcT\rangle\rangle}$ and
let $\{ {\mathcal I}_{\alpha}\}$, $\alpha\in A$, be the
collection of connected components of the timelike boundary
$\pi^{-1}(\mcT )$. Let us define ${\langle\langle
I\rangle\rangle}_{\alpha,\beta}:= I^+({\mathcal
I}_{\alpha})\cap I^-({\mathcal I}_{\beta})$. We claim that the
collection of sets ${\langle\langle
I\rangle\rangle}_{\alpha,\beta}$ forms an open cover of $M$.
Indeed, let $p\in M$; since $\pi(p)\in
{\langle\langle\mcT\rangle\rangle}$ there exists a causal curve
through $\pi(p)$ which starts and ends in $\mcT$. Then $\gamma$
lifts to a causal curve through $p$ which starts in some
${\mathcal I}_\alpha$ and ends in some ${\mathcal I}_\beta$,
hence the result. Now, by Theorem~\ref{topcenwb} the sets
$I^+({\mathcal I}_{\alpha})\cap I^-({\mathcal I}_{\beta})$ are
empty if $\alpha\neq\beta$. It follows that the sets
${\langle\langle I\rangle\rangle}_{\alpha,\alpha}$ are pairwise
disjoint, cover $M$, and since $M$ is connected we conclude
that $\vert A\vert =1$ and hence $\pi^{-1}(\mcT )$ is
connected. The result now follows from the following
topological result~\cite[Lemma~3.2]{GSWW}:
\end{proof}

\begin{Proposition}\label{last}
Let $M$ and $S$ be topological manifolds, $\imath\colon
S\hookrightarrow M$ an embedding and $\pi\colon M^*\to M$ the
universal cover of $M$. If $\pi^{-1}(S)$ is connected then the
induced group homomorphism $\imath_*\colon \pi_1(S)\to \pi_1(M)$
is surjective.
\end{Proposition}

\medskip

\section{Kaluza-Klein asymptotics}
 \label{sKKa}
\renewcommand{\KK}{Q}

In the sections that follow we shall apply
Theorem~\ref{topcenwb} to obtain topological information about
space-times with \emph{Kaluza-Klein} asymptotics:   we shall
say that $\Sext$ is a \emph{Kaluza-Klein asymptotic end}, or
\emph{asymptotic end} for short, if $\Sext$ is diffeomorphic to
$ \R_r\times N \times \KK$, where $N$ and $\KK$ are compact
manifolds. The notation $\R_r$ is meant to convey the
information that we  denote by $r$ the coordinate running along
an $\R$ factor. Let $\mathring m_r$ be a family of Riemannian
metrics paremeterized by $r$, let $\mathring k$ be a fixed
Riemannian metric on $\KK$, let finally $\mathring \lambda$ and
$\mathring\nu$ be two functions on $\R$, the \emph{reference
metric} $\mathring g$ on $\R_t\times \Sext$ is defined as
\bel{basemetdef}
 \mathring g = -e^{2\mathring \lambda(r)}dt^2 + \underbrace{e^{-2\mathring \nu(r)}dr^2 + \mathring m_r+\mathring k}_{=: \mathring \threeg}
 \;.
\ee
The reason for treating $N$ and $\KK$ separately is that the
metrics $\mmr$ are allowed to depend on $r$ (in the examples
below we will actually have $\mmr=r^2 \mathring m$, for a fixed metric
$\mathring m$), while $\mk$ is not. The manifold $\R_t\times
\R_r \times N$ will be referred to as the \emph{base manifold},
while $\KK$ can be thought of as the internal space of
Kaluza-Klein theory (see, e.g.,~\cite{Coquereaux:1988ne}).

To apply our previous results, we will need the hypothesis that
the hypersurfaces
$$
 \mcT_R:=\{r=R\}
$$
are inner future and past trapped for the reference metric
$\mg$. We define the \emph{outwards} pointing $\mg$--normal to
$\mcT_R$ to be $n:=e^{\mnju}\partial_r$, and the two null
future normals $n_\pm$ to $\{t=\const\;,\ r=\const'\}$ are
given by $n_\pm = e^{-\mathring \lambda}\partial_t \pm n$. The
requirement of ``mean outwards null $\mg$--convexity" of $\mcT_R$
reads
\bel{mtpm}
 \pm \mathring \theta_\pm = \frac {e^{\mathring \nu - \mathring \lambda}} {\sqrt{ \det \mmr}} \partial_r(\sqrt{ \det \mmr} e^{\mathring \lambda}) >0
 \;.
\ee
We will be interested in metrics $g$ which are asymptotic, as
$r$ goes to infinity, to metrics of the above form. The
convergence of $g$ to $\mathring g$ should be such that the
positivity of $\pm\theta_\pm$ holds, for $R$ large enough,
uniformly over compact sets of the $t$ variable. Two special
cases seem to be of particular interest, with asymptotically
flat, or asymptotically anti-de Sitter base metrics.

 \subsection{Asymptotically flat base manifolds}
\label{ssAfbm}
A special case of the above arises when
$\Sext$ is
diffeomorphic to $\left(\R^n\setminus \overline B(R)\right)
\times \KK$, where $\overline B(R)$ is a closed coordinate ball
of radius $R$, thus the manifold $N$ is an $(n-1)$--dimensional sphere.
In dimension $n\ge 3$ we take  $\mathring g = -dt^2\oplus\mathring
\threeg$, where $\mathring\threeg=\delta \oplus \mathring k$,
and where $\delta $ is the Euclidean metric on $\R^n$. If $n=2$, in \eq{basemetdef}
we take $\mathring \lambda = \mathring \nu=0$ and
$\mmr=r^2 d\varphi^2$, where $\varphi$ is a coordinate on $S^1$ which does not  \emph{necessarily}  range
over $[0,2\pi]$.  Thus, for all $n\ge 2$  we have
$\mathring \lambda = \mathring \nu=0$ and $\mathring m_r = r^2
d\Omega^2$, where $d\Omega^2$ is the round metric on
$S^{n-1}$; strictly speaking, $\R_r$ is then $(R,\infty)$, a
set diffeomorphic to $\R$.
Metrics
$g$ which asymptote to this $\mg$ as $r$ tends to infinity will be
said to have an \emph{asymptotically flat base manifold}.
 \Eq{mtpm} gives
\bel{mtpm2}
 \pm \mathring \theta_\pm = \frac {n-1} {r}
\ee
which is positive, as required.

We shall say that a Riemannian metric $\threeg$ on $\Sext$ is
 \emph{Kaluza-Klein asymptotically flat}, or
 \emph{$KK$--asymptotically flat} for short, if there exists
 $\alpha>0$ and $k\ge 1$ such that for $0\le \ell \le k$ we
\bel{meddec} \zD_{i_1}\ldots \zD_{i_\ell} (\threeg-\mathring \threeg) = O(r^{-\alpha-\ell})
 \;, \ee
where $\zD$ denotes the Levi-Civita connection of
$\mathring \threeg$, and $r$ is the radius function in $\R^n$,
$r:=\sqrt{(x^1)^2+\ldots (x^n)^2}$, with the $x^i$'s being any
Euclidean coordinates of $(\R^n,\delta)$.   We shall say that a
general relativistic initial data set $(\Sext,\threeg,K) $   is
\emph{Kaluza-Klein asymptotically flat}, or
\emph{$KK$--asymptotically flat}, if $(\Sext,\threeg)$ is
$KK$--asymptotically flat and if for  $0\le \ell \le k-1$ we
have
\bel{Kdec} \zD_{i_1}\ldots \zD_{i_\ell} K =
O(r^{-\alpha-1-\ell})
 \;.
\ee

A space-time  $(\mcM,\fourg)$  will be said to contain a
\emph{Kaluza-Klein asymptotically flat region} if there exists
a subset of $\mcM$, denoted by $\Mext$, and a time function $t$
on $\Mext$, such that the initial data $(g,K)$ induced by
$\fourg$ on the level sets of $t$   are $KK$--asymptotically
flat.

All this reduces to the usual notion of asymptotic flatness
when $\KK$ is the manifold consisting of a single point; a
similar comment applies to the next section.

Let $\mcT_R=\{r=R\}$ be a level set of $r$ in $\Mext$.  Then the unit outwards pointing conormal
$n^\flat$ to $\mcT_R $ is $n^\flat=
(1+O(r^{-\alpha}))dr$. This implies that the future directed
null vector fields normal to the foliation of $\mcT_R$ by
the level sets of $t$ take the form $n^\pm =
\partial_t\pm \frac {x^i}{r}\partial_i +O(r^{-\alpha})$,
leading to (compare~\eq{mtpm2})
\bel{trappingT}
 \pm \theta^\pm =\frac{n-1}r +O(r^{-\alpha-1}) >0 \ \mbox{for $r$ large enough}
 \;.
\ee

\subsection{Asymptotically anti-de Sitter base manifolds}
 \label{ssAadS}
\newcommand{\tg}{\tilde g}

We consider, now, manifolds with asymptotically anti-de Sitter
base metrics. The  base reference metric is taken of the form
\bel{basemet}
 -e^{2\mathring \lambda(r)}dt^2 +  { e^{-2\mathring \nu(r)}dr^2 + \mathring m_r}
 \;, \qquad \mathring m_r = r^2
\mathring m\;,
\ee
which can be thought of as being a generalised Kottler metric,
where $\mathring m$ is an Einstein metric on the compact
$(n-1)$--dimensional manifold $N$, $n\ge 2$. Furthermore,
$$
 e^{2\mathring \lambda(r)}=e^{ 2\mathring \nu(r)} = \mathring\alpha r^2 + \mathring\beta\;,
$$
for some suitable constants $\mathring\alpha>0$ and $\mathring\beta \in \R$,
which can be chosen so that \eq{basemet} is an Einstein metric:
Indeed, if $\KK$ has dimension $k$,
then  $\mathring g$ will  satisfy the vacuum Einstein equations with
cosmological constant $\Lambda$ if  $\mk$ is an Einstein metric
with scalar curvature $2k\Lambda/(n+k-1)$, while $\mathring\alpha=
-2\Lambda/n(n+k-1)$, and $\mathring\beta = R(\mathring m)/(n-1)(n-2)$ for $n>2$,
while $\mathring \beta$ is arbitrary if $n=2$,
where $R(\mathring m)$ is the scalar curvature of the metric
$\mathring m$ (compare~\cite{CadeauWoolgar,BTZ}).

In a manner somewhat analogous to the previous section,  with
 decay requirements adapted to the problem at hand, we shall
 say that a Riemannian metric $\threeg$ on $\Sext$ is
 \emph{$KK$--asymptotically adS}  if there exist a real number $\alpha>1$
 and an integer $k\ge 1$ such that for $0\le \ell \le k$ we have
\bel{meddec2} |\zD_{i_1}\ldots \zD_{i_\ell} (\threeg-\mathring \threeg)|_{\mathring \threeg} = O(r^{-\alpha})
 \;, \ee
 where $|\cdot |_{\mathring \threeg}$ is the norm of a tensor
with respect to $\mathring \threeg$, and $r$ is a ``radial
coordinate" as in \eq{basemet}.   We shall say that a general
relativistic initial data set $(\Sext,\threeg,K) $   is
\emph{$KK$--asymptotically adS}, if $(\Sext,\threeg)$ is
$KK$--asymptotically adS  and if for  $0\le \ell \le k-1$ we
have
\bel{Kdec2} |\zD_{i_1}\ldots \zD_{i_\ell} K |_{\mathring \threeg}=
O(r^{-\alpha})
 \;.
\ee
Finally, a space-time  $(\mcM,\fourg)$  will be said to contain a
\emph{Kaluza-Klein asymptotically adS region} if there exists
a subset of $\mcM$, denoted by $\Mext$, and a time function $t$
on $\Mext$, such that the initial data $(g,K)$ induced by
$\fourg$ on the level sets of $t$   are $KK$--asymptotically
adS.

The fact that $KK$--asymptotically adS metrics have the right
null convexity properties is easiest to see using the conformal
compactifiability properties of the base metric. Suppose, to
start with, that $\KK$ consists only of one point, so that
$\mathring k=0$. Suppose further that $g$ has a conformal
compactification in the usual Penrose sense, so that the
unphysical metric  $\tg_{\mu\nu}=\Omega^{-2}g_{\mu\nu}$ extends
smoothly to a conformal boundary at which $\Omega$ vanishes;
this is certainly the case for the reference metrics $\mg$ of
\eq{basemet}, and will also hold for a large class of
asymptotically adS metrics as defined above. We define the
outwards directed $\tg$--unit normal to the level sets of
$\Omega$ to be
$$
 \tilde n^\mu = -\frac{\tg^{\mu\nu}\partial_\nu \Omega}{\sqrt{ \tg^{\alpha\beta}\partial_\alpha \Omega\partial_\beta \Omega }}
$$
(the minus sign being justified by the fact that $\Omega$
decreases as the conformal boundary $\{\Omega =0\}$ is
approached). Let, finally, $t$ be a time function on the
conformally completed manifold $\bmcM$ such that the $\tg$-unit
timelike vector field $\tilde T$  normal to the level sets of
$t$ is tangent to the conformal boundary; thus $\tilde
T(\Omega) = \Omega \psi$ for some function $\psi$ which is
smooth on $\bmcM$. Then $n^\mu=\Omega \tilde n^\mu$ and $T^\mu
= \Omega \tilde T^\mu$ are unit and normal to $\{t=\const,
\Omega = \const'\}$. So, in space-time dimension $n+1$,
\beal{trpads}
\pm \theta_\pm  & = &  \nabla_\mu (\pm T^\mu + n^\mu)
 \\ \nonumber
 & = & \frac 1 {\sqrt {|\det g|}} \partial_\mu \left(  \sqrt {|\det g|} (\pm T^\mu + n^\mu) \right)
 \\  \nonumber
 & = & \frac {\Omega^{n+1}} {\sqrt {|\det \tg|}} \partial_\mu \left( \Omega^{-n-1} \sqrt {|\det \tg|} \Omega (\pm \tilde T^\mu + \tilde n^\mu) \right)
 \\
 & = &
 n |d \Omega|_{\tg} + O(\Omega)
 \;,
 \nonumber
\eea
which is positive for $\Omega$ small enough (in the last
equation $n$ is the space-dimension, not to be confused with
the unit normal to the level sets of $\Omega$). It is now a
simple exercise to check that, for $KK$--asymptotically adS
metrics, the correction terms arising from $\mk$, and from the
error terms in \eq{meddec2}-\eq{Kdec2} will not affect
positivity of $\pm \theta_\pm$ whenever $\alpha>1$, as required
above.

\subsection{Uniform $KK$--asymptotic ends}

We shall say that a $KK$--asymptotically flat region, or a
$KK$--asymptotically adS region, is \emph{uniform of order $k$}
if there exists a time function $t$ such that the estimates
\eq{meddec}-\eq{Kdec}, or \eq{meddec2}-\eq{Kdec2}, hold  with
constants independent of $t$.

\section{Topological censorship for uniform $KK$--asymptotic ends}
 \label{sghsc}

In this section we shall consider manifolds with
 $KK$--asymptotically flat or $KK$--asymptotically adS regions.
 Now, our approach  to topological censorship in this work
 requires uniformity in time of the mean null extrinsic
 curvatures of the spheres $\{t=\const\;,\ r=\const'\}$. This
 might conceivably hold for a wide class of dynamical metrics,
 but how large is the corresponding class of metrics remains to
 be seen.  Now, the applications we  have in
 mind for our results~\cite{ChHighDim} concern stationary
 metrics, in which case the uniformity is easy to guarantee by
 an obvious choice of time functions. Hence the uniformity
 hypothesis is quite reasonable from this perspective.

 Consider, first, a space-time with
a Killing vector field $X$, with complete orbits, containing a
 $KK$--asymptotic end $\Sext$. Then $X$ will be called
 \emph{stationary} if $X$ is timelike on $\Sext$ and
 approaches, as $r$ goes to infinity, $\partial_t$ in the coordinate system of \eq{basemetdef};%
 \footnote{For metrics which are asymptotically flat in the
 usual (rather than $KK$) sense, the existence of such
 coordinates   can be established for  Killing vectors which
 are timelike on $\Sext$,   whenever the initial data set
 satisfies the conditions of the positive energy theorem. It is
 likely that a similar result holds for $KK$--asymptotically
 flat or adS metrics, but we have not investigated this issue
 any further.}
 $(\mcM,\fourg)$ will then be called stationary.
Similarly to the standard asymptotically flat case, we set %
$$
 \Mext:= \cup_{t\in \R} \phi_t[X](\Sext)\;,
$$
where  $\phi_t[X]$ denotes the flow of $X$. Assuming
stationarity, the \emph{domain of outer communications} is
defined as in~\cite{ChWald1,ChCo}:
\bel{docdef}
 \doc:= I^-(\Mext)\cap I^+(\Mext)
 \;.
\ee
More generally, let $(\mcM,\fourg)$ admit a time function $t$
ranging over an open interval $I$ (not necessarily equal to
$\R$),  and   a radius function $r$ as in \eq{basemetdef},
with $KK$--asymptotic level sets which are \emph{uniform of
order zero}. We then set
$$
 \Mext:= \{p\in \mcM:\ r(p)\ge R_0\}
$$
for some $R_0$ chosen large enough so that for any $R \ge R_0$ we have
\bel{equdoc}
 J^\pm(\Mext )=J^\pm(\{r=R \})
 \;.
\ee
To see that such an $R_0$ exists, note that the inclusion
 $J^\pm(\Mext )\supset J^\pm(\{r=R \})$ is obvious whenever
 $\{r=R\}\subset \Mext$. To justify the opposite inclusion let,
 say, $x\in J^-(\Mext)$, so there exists a future directed
 causal curve from $x$ to some point $(t,p)\in \Mext$, thus
 $p\in \Sext$. We need to show that there exists a future
 directed causal curve from $(t,p)$ to a point $(t',q)\in
 \{r=R\}$. This follows from the somewhat more general fact,
 that for any $t$ and for any two points   $p,q\in\Sext$ such
  that $r(p)\ge R_0$ and $r(q)\ge R_0$, there exists a causal
  curve $\gamma(s)=(t+ \alpha s, \sigma(s))$ such that
  $\sigma(0)=p$ and $\sigma(0)=q$, with $\alpha$ and $\sigma$
  independent of $t$. Now, existence of $\sigma$ follows from
  connectedness of $\Sext$. Next, the existence of a
  $t$--independent (large) constant $\alpha$ so that $\gamma$
  is causal for $\mathring g$ follows immediately from the
  form of the metric $\mathring g$. Finally, it should be clear
  from   uniformity in time of the error terms   that,
  increasing  $\alpha$ if necessary, $\gamma$ will also be
  causal for $g$, independently of $t$, provided $R_0$ is
  chosen large enough.

The domain of outer communications is again defined by \eq{docdef}.

If the asymptotic estimates are moreover uniform to order one,
we choose $R_0$ large enough so that all level sets of
$\{r=R\}$, with $R$ sufficiently large are \emph{future and
past inner trapped}.

\begin{Remark}
{\rm As shown in Appendix~\ref{SALRL}, there exist vacuum
space-times which are uniformly asymptotically flat  to order
zero, and for which the null convexity conditions needed for
our arguments are satisfied even though the asymptotic flatness
estimates \eq{meddec}-\eq{Kdec} are \emph{not} uniform to order
one. For simplicity, in this section we shall only formulate
our theorems assuming uniformity to order one, but it should be
clear to the reader that the results hold e.g. for metrics with
the asymptotic behavior as in Appendix~\ref{SALRL}.
}
\end{Remark}

\medskip

Let us consider a space-time $(\mcM,\fourg)$ with several
$KK$--asymptotic  regions $\Mext^\lambda$, $\lambda
\in\Lambda$, each generating its own domain of outer
communications. We assume that all regions are uniform to order
one with respect to a Cauchy time function $t$. Let
$\mcT_\lambda\subset \Mext^\lambda$ be defined as $\{r=\hat
R_\lambda\}$ for an appropriately large $R_\lambda$. Consider
the manifold obtained by removing from the original space-time
the asymptotic regions $ \{r> \hat R_\lambda \}\subset
\Mext^\lambda$; this is a manifold with boundary $\mcT=\cup_\lambda \mcT_\lambda$, each
connected component $\mcT_\lambda$ being both future   and past
inwards trapped. From \eq{equdoc} we have
$J^\pm(\Mext^\lambda)=J^\pm(\mcT_\lambda)$. Then the following
result is a  straightforward consequence of Theorem
\ref{TGconv}:

\begin{Theorem}
\label{topocensor} If $(\mcM,\fourg)$ is a globally hyperbolic with  $KK$--asymptotic ends, uniform to order one,
 satisfying
the null energy condition \eq{NEC}, then  \bel{asg2}
 J^+(\mcMext^{\lambda_1}) \cap J^{-}( \mcMext^{\lambda_2})=\emptyset \
 \text{ whenever  }\  \Mext^ {\lambda_1}\cap \Mext^{\lambda_2}=\emptyset\,.
\ee
\end{Theorem}

Next, Theorems~\ref{topcenwb} and \ref{stopcen} yield the
following result on topological censorship for stationary
$KK$--asymptotically flat spacetimes:

\begin{Theorem}
Let $(\mcM,\fourg)$ be a space-time satisfying the null energy
condition, and containing a   $KK$--asymptotic  end $\Mext$,
uniform to order one. Suppose further that $\doc$ is globally
hyperbolic. Then every causal curve in $\doc$ with endpoints in
$\Mext$ is homotopic to a curve in $\Mext$. Moreover the map
$j_*\colon \pi_1(\Mext)\to \pi_1(\doc)$ is surjective.
\end{Theorem}

\begin{proof}
 It suffices to prove the second statement. Let $\hat{R}>R$ and
 $\mcT=\{r=\hat{R}\}$ be defined as in the previous result. Let
 $\alpha$ be a loop in $\doc$ based at $p_0$, and let $c$ be
 the radial curve from $p_0$ to $p\in\mcT$. Then since
 $\doc=\langle\langle \mcT\rangle\rangle$, by
 Theorem~\ref{stopcen} the loop $c*\alpha *c^{-}$, where $c^-$ denotes $c$ followed backwards,
is homotopic to a loop $\beta$ in $\mcT$ based at $p$. Thus
 $\alpha$ is in turn homotopic to $c^{-}*\beta*c$, which is a
 loop that lies entirely in $\Mext$ hence establishing the
 result.
 \end{proof}

For future reference, we point out the following special case
of Proposition~\ref{Pnotrapped0}, which follows immediately
from the fact that large level sets of $r$ are inner trapped:

\begin{Proposition}
 \label{Pnotrapped}
Let $(\mcM,\fourg)$ be a stationary, asymptotically flat, or $KK$--asymptotically flat   globally
 hyperbolic space-time satisfying the null energy condition.
 Let $S\subset \doc$ be future inwards marginally trapped.  There exists a
 large constant $R_1$ such that for all $R_2 \ge R_1$ there are
 no future directed null geodesics starting inwardly at $S$,
 ending inwardly at
 $\{r=R_2\}\subset \Mext$, and locally minimising the time of flight.
\end{Proposition}

\medskip

Now we proceed to prove the main theorem for quotients of $KK$--asymptotically flat spacetimes.

\begin{Theorem}
\label{TCdsc} Let $(\mcM,\fourg)$ be a space-time satisfying
the null energy condition, and containing a
$KK$--asymptotically flat region, or a $KK$--asymptotically adS
region, with the asymptotic estimates uniform to order one.
Suppose that $\doc$ is globally hyperbolic, and that there
exists an action of a group $  G_s$
 on $\doc$ by isometries which, on $\Mext\approx \R\times
 \Sext$, takes  the form
$$
       g\cdot (t,p)= (t,g\cdot p)
 \;.
$$
If $\Sext/G_s$ simply connected, then so
is $\doc/G_s$.
\end{Theorem}

\begin{Remark} {\rm
 A variation on the proof below, using an exhaustion argument,
 shows that the result remains valid if the asymptotic decay
 estimates are uniform in $t$ to order zero, and uniform over
 compact sets in $t$ to order one. In this case the
 hypersurfaces $\{r=R\}$ are not necessarily trapped, but there
 exists a sequence $R_k$ such that the hypersurfaces
 $\{r=R_k,|t|<k\}$ are.
 }
\end{Remark}

\begin{proof}
If the action of $G_s$ is such that the projection
$\doc\to\doc/G_s$ has the homotopy lifting property (see,
e.g., \cite{Hatcher}), then the following argument applies:
Consider the  commutative diagram
$$\begin{matrix}
\Mext & \stackrel{i}{\longrightarrow} & \doc \\
q\downarrow & & \downarrow p\\
\Mext /G_s & \stackrel{j}{\longrightarrow} & \doc /G_s
\end{matrix}
$$
where $p$ and $q$ are the standard projections, $i$ the standard inclusion and $j$ the map induced by $i$. Thus we have the corresponding commutative diagram
$$\begin{matrix}
\pi_1(\Mext) & \stackrel{i_*}{\longrightarrow} & \pi_1(\doc )\\
q_*\downarrow & & \downarrow p_*\\
\pi_1(\Mext /G_s) & \stackrel{j_*}{\longrightarrow} & \pi_1(\doc /G_s)
\end{matrix}
$$
of fundamental groups. By Theorem~\ref{topocensor},  $i_*$
is onto.  Finally notice that $p_*$ and $q_*$ are onto since
$p$ and $q$ have the homotopy lifting property. Hence
$j_*$ is onto and as a consequence $\doc /G_s$ is simply
connected if $\Mext /G_s=\R\times (\Sext/G_s)$ is.

{The homotopy lifting property of the action is known to hold in many significant cases (e.g., when  the  action is free),}
 but it is not clear  whether it holds in sufficient generality. However, one can
proceed as follows: Let $\pi$ denote the projection map
$$
 \pi:\mcM\to \mcM/G_s
 \;.
$$
We start by constructing a  covering space, $\hmcM$, of $\mcM$:
Choose $p\in \mcM$ and let $\Omega$ be the set of {continuous}
 paths in $\mcM$ starting at
$p$. We shall say that the paths $\gamma_a\in \Omega$, $a=1,2$,
are equivalent, writing $\gamma_1\sim \gamma_2$, if they share
their end point, and if the projection $\pi(\gamma_1 *
\gamma _2^-)$  of the path $ \gamma_1 *  \gamma
_2^-$,
  obtained by concatenating $\gamma_1$ with $\gamma_2$
followed backwards, is homotopically trivial in $\mcM/G_s$. We
set
$$
 \hmcM:=\Omega/\sim
 \;.
$$
By the usual arguments (see, e.g., the proof
of~\cite[Theorem~12.8]{Lee:TM}) $\hmcM$ is a topological
covering of $\mcM$, while~\cite[Proposition 2.12]{Lee:SM} shows
that $\hmcM$ is a smooth manifold. (In fact, $\hmcM$ is the covering space of $\mcM$ associated
with the subgroup  Ker$\, \pi_* \subset \pi_1(M)$.)
The covering is
trivial if and only if $\mcM/G_s$ is simply connected.

  Since $\Sext/G_s$ is simply connected, the quotient
$\Mext/G_s=\R\times (\Sext/G_s)$ also is, which implies that
$\pi^{-1}(\Mext)\subset \hmcM$ is the union of pairwise
disjoint diffeomorphic copies $\Mext^\lambda$,
$\lambda\in\Lambda$,  of $\Mext$, for some index set $\Lambda$.
Each $\Mext^\lambda$ comes with an associated open domain of
dependence $\docl\subset \hmcM$. As in the proof of
Theorem~\ref{stopcen}, the $\docl$'s form an open cover of
$\hmcM$. Moreover, by Theorem~\ref{topocensor} they are
pairwise disjoint. Connectedness of $\hmcM$ implies that
$\Lambda$ is a singleton $\{\lambda_*\}$, with $\hmcM=\docls$,
hence $\hmcM=\mcM$, and the result follows.
\end{proof}

\subsection{Existence of twist potentials}
 \label{ssetp}
 We
  turn now our attention to the question of existence of
  twist potentials. The problem is the
  following: suppose that $\omega$ is a closed one form on a
  domain of outer communications $\doc$. For $i=1,\ldots r$ let
$X_i$ be the basis of a Lie algebra of Killing vectors
generating a connected group $G$ of isometries and suppose that
\bel{omeinvi}
\forall i \qquad
 \mcL_{X_i}\omega = 0=\omega(X_i)
 \;.
\ee
If $\doc$ is simply connected, then there exists a
$G$-invariant function $v$ such that $\omega=dv$. More
generally, if $\doc/G$ is a simply connected \emph{manifold},
then $\omega$ descends to a closed one-form on $\doc/G$, and
again existence of the potential $v$ follows. Let us show that
the hypothesis that $\doc/G$ is a manifold  can be replaced by
the weaker condition, that the projection map $\doc\to \doc/G$
has the \emph{path homotopy lifting property}, namely: every
homotopy of paths in $\doc/G$ can be lifted to a continuous
family of paths in $\doc$:

\begin{Proposition}
 \label{Ptwist} If $\doc/G$ is  simply connected, and if the path
 homotopy lifting property holds, then there exists a
 $G$--invariant function $v$ on $\doc$ so that $\omega=dv$.
\end{Proposition}

\proof To simplify notations, let $\mcM=\doc$ with the induced
metric. Choose a point $p\in \mcM$, let $\gamma:[0,1]\to \mcM$
be any path with $\gamma(0)=p$, set
$$
 v_\gamma=\int_\gamma \omega
 \;,
$$
we need to show that $v_\gamma = 0$ whenever
$\gamma(0)=\gamma(1)$. Let $\mathring \gamma^\flat$ be the projection  to
$\mcM/G$ of a loop $\mathring \gamma$ through $p$, since $\mcM/G$ is
simply connected there exists a continuous one-parameter family
of paths $\gamma^\flat_t$, $t\in [0,1]$, so that
$\gamma^\flat_0=\mathring \gamma^\flat$,
$\gamma^\flat_t(0)=\gamma^\flat_t(1)=\mathring\gamma^\flat(0)$,
$\gamma^\flat_1(s)=\mathring \gamma^\flat(0)$. Let $\gamma_t$ be
any continuous lift of $\gamma^\flat_t$ to $\mcM$ which is also
continuous in $t$, such that $\gamma_t(1)=p$.  Then
$\gamma_t(0)=g_t p$ for some continuous $g_t \in G$. We can thus obtain a closed path through $p$, denoted
by $\hat \gamma_t$, by following $\gamma_t$ from $p$ to $\gamma_t(0)$,
and then following the path
$$
 [0,t]\ni s\mapsto g_{t-s}p
\;.
$$
Since
$\gamma_1$ is trivial, so is $\hat \gamma_1=\gamma_1$, so that  $v_{\hat \gamma_1}=0$. The family $\hat \gamma_t$
provides thus a homotopy of $\hat \gamma_0$ with $\hat \gamma_1$, and by homotopy invariance
$$
0= v_{\gamma_1} = v_{\hat \gamma_1} = v_{\hat \gamma_0}= v_{ \gamma_0}
 \;.
$$
Next, using the fact that both $\gamma_0$ and $\mathring \gamma$ project to
$\mathring \gamma^\flat$, we will show that
\bel{godeq}
 v_{\gamma_0} = v_{\mathring \gamma}
 \;,
\ee
which will establish the result.

Let $s\in [0,1]$,  set $r:=\mathring\gamma(s)$, let
$\mcO_r\subset \mcO$ denote any sufficiently small simply
connected neighborhood of $r$,  and let $v_r$ denote the
solution on $\mcO_r$ of
\bel{vpsol}
 dv_r=\omega\;,\quad  v_r(r)=0
 \;.
\ee
Let $\mcU_r=G \mcO_r$ be the orbit of $G$ through $\mcO_r$, for
$p'\in \mcU_p$  there exists $\hat p\in \mcO_r$ and $g\in G$
such that $p'=g\hat p$. Set $v_r(p'):=v_r(\hat p)$, this is
well defined as the right-hand-side is independent of the
choice of $g$ and $q$ by \eq{omeinvi}. Then $v_r$ is a solution
of \eq{vpsol} on $\mcU_r$, and for all $s$ such that $\mathring
\gamma(s)\in \mcU_r$ we have
$$
v_r(\mathring \gamma(s))= v_r ( \gamma_0(s))
 \;.
$$
It follows that for any interval $[s_1,s_2]$ such that
$\mathring \gamma([s_1,s_2])\subset \mcO_r$ we have
$$
 \int_{\mathring \gamma([s_1,s_2])} \omega = v_r(\mathring \gamma(s_2))-v_r(\mathring \gamma(s_1))=
 v_r(\gamma_0(s_2))- v_r(\gamma_0(s_1))=
 \int_{ \gamma_0([s_1,s_2])} \omega
 \;.
$$
A covering argument finishes the proof.
\qed
\section{Weakly future trapped surfaces are invisible}
 \label{swftsi}

Yet another application of the ideas above is the following result,
which is part of folklore knowledge in general relativity,
without a satisfactory proof available elsewhere in the
literature:

\newcommand{\bmcD}{\,\,\widetilde{\!\!\mcD}}%
\begin{Theorem}\label{notrapped0} Let $(\mcM,\fourg)$ be an
 asymptotically flat spacetime, in the sense of admitting a
 regular future conformal completion $\bmcM = \mcM \cup \scri^+$,
 where $\scri^+$ is a connected null hypersurface, such that,
 \begin{enumerate}
 \item $\bmcD = \mcD \cup \scri^+$ is globally hyperbolic, where $\mcD = I^-(\scri^+,\bmcM)$, and
 \item  for any compact set $K \subset \mcD$, $J^+(K, \bmcD)$ does not
 contain all of~$\scri^+$ (``$i^0$-avoidance").
  \end{enumerate}
If the  NEC holds on $\mcD$, then there are no compact future weakly trapped
submanifolds within $\mcD$.
\end{Theorem}

\begin{Remarks}
 \label{Rglobhyp}
 {\rm
 \begin{enumerate}
 \item
 Note that if $\mcM\cup\scri^+$ is globally hyperbolic, then $\bmcD$ also is.
 \item Compare~\cite[Appendix~B]{ChDGH} for a discussion of
     issues that arise in a related context.
 \end{enumerate}
 }
\end{Remarks}

\begin{proof} We begin by noting that the global hyperbolicity of $\bmcD$ implies
that $\bmcD$ is causally simple, i.e., that sets of the form
$J^+(K,\bmcD)$ are closed
in $\bmcD$ for all compact sets $K$.   Suppose $S$ is a compact future weakly trapped
submanifold in $\mcD$.
Let $q$ be a point on  $\del(J^+(S,\bmcD) \cap \scri^+)  = \dot
J^+(S,\bmcD) \cap \scri^+ $, which is nonempty by
$i^0$-avoidance.
  Since $\dot J^+(S,\bmcD) = J^+(S,\bmcD)
\setminus I^+(S,\bmcD)$, there exists an achronal null geodesic
$\ga: [a,b] \to \bmcD$, with $\ga(a) \in S$ and $\ga(b) = q$,
emanating orthogonally from $S$, without $S$-conjugate points
on $[a,b)$. In particular, $\dot J^+(S)$ is a smooth null
hypersurface near $\gamma{([a,b))}$.  Below we show that for a
suitably chosen point $q \in \dot J^+(S,\bmcD) \cap \scri^+$,
there exists a spacelike hypersurface $S_+$ in $\scri^+$ that
passes through $q$ and does not meet $I^+(S,\bmcD)$.   Given
this, the proof may now be completed along the lines of the
proof of Proposition \ref{Pnotrapped0}.    Since $S_+$ does not
meet $I^+(S,\bmcD)$, one easily argues that   $\dot J^-(S_+)$
is a smooth null hypersurface near $S_+$ that contains  a
segment $\gamma([b-\varepsilon,b])$ and  lies to the causal
past of $\dot J^+(S)$.

Let $\tilde K$ be a future directed outward pointing null
vector at $q$ orthogonal to $S_+$ in the unphysical metric
$\tilde \fourg = \Omega^2 \fourg$. Since $\Omega$ decreases to
the future along $\ga$ near $q$, we can choose $\tilde K$ so
that $ \tilde K(\Omega) =\tilde\fourg(\tilde K, \tilde\nabla
\Omega) = -1$. Now extend $\tilde K$ to a null vector field
tangent to $\dot J^-(S_+)$ near $q$, and let $K = \Omega \tilde
K$.  A computation, using basic properties of  conformal
transformations, shows that the divergence $\theta$ of $\dot
J^-(S_+)$ with respect to $K$ in the physical metric $\fourg$
is related to the divergence $\tilde\theta$ of $\dot J^-(S_+)$
with respect to $\tilde K$ in the unphysical metric
$\tilde\fourg$ by, in space-time dimension $n+1$,
$$
\theta = -(n-1) \tilde K(\Omega) + \Omega\, \tilde \theta \,.
$$
It follows that $\dot J^-(S_+)$ will have positive null divergence at
points of $\ga$ close to $q$.  On the other hand, as in the proof of
Proposition \ref{Pnotrapped0}, $\dot J^+(S)$, has nonpositive
null divergence along $\ga$, and we are again led to a
contradiction of the maximum principle for null hypersurfaces.

We conclude the proof by explaining how to choose  $q$ and $S_+$.
For this purpose we introduce a Riemannian metric on $\scri^+$,
with respect to which the following constructions  are carried
out. Fix $q_0 \in  \dot J^+(S,\bmcD) \cap \scri^+$, and let $U
\subset \scri^+$ be a convex normal neighborhood of $q_0$.  By
choosing a point $p \in U$, $p  \notin J^+(S,\bmcD)$,
sufficiently close to $q_0$, we obtain a point $q \in  \dot
J^+(S,\bmcD) \cap \scri^+$, such that the geodesic segment
$\overline{pq}$ in $U$ realizes the distance from $p$ to $\dot
J^+(S,\bmcD) \cap \scri^+$.   Now
 let $S_+$ be the distance sphere in $U$ centered at $p$ and
 passing through $q$.   $S_+$ is a smooth hypersurface in
 $\scri^+$ that does not meet $I^+(S,\bmcD)$.  It follows that
 $S_+$  intersects the generator of  $\scri^+$ through $q$
 transversely, and hence  is spacelike near $q$.
 To see this,
 let $\ga$ be the null geodesic from $S$ to $q$ as in the preceding
 paragraph.  For  $x \in \ga$ sufficiently close to $q$,  $S' =
  \dot J^+(x,\bmcD) \cap \scri^+$ will be, in the vicinity of $q$,  a smooth hypersurface
  in $\scri^+$ transverse to the null generator of $\scri^+$
  through~$q$.  But, since $S' \subset J^+(S,\bmcD)$,
  $S_+$ must meet $S'$ tangentially at $q$.
  Hence $q$ is the desired point in $\dot J^+(S,\bmcD) \cap
 \scri^+$ and $S_+$,  suitably restricted, is the desired
 spacelike hypersurface in $\scri^+$.
 \end{proof}

\begin{Remark}
{\rm
An entirely analogous result holds for asymptotically anti-de
Sitter spacetimes, in the sense of admitting a regular
conformal completion, with timelike conformal infinity $\scri$,
and can be proved in a similar fashion.
}
\end{Remark}

We further note that Proposition~\ref{Pnotrapped0} may be used
to obtain a version of Theorem \ref{notrapped0} for spacetimes
$(\mcM,\fourg)$ with   $KK$--asymptotic ends, as follows.

\begin{Theorem}\label{notrapped} Let $(\mcM,\fourg)$ be a
$KK$--asymptotically flat or $KK$--asymptotically anti-de
Sitter space-time with the asymptotic estimates uniform to
order one. If $(\mcM,\fourg)$ contains a globally hyperbolic
domain of outer communications $\doc$ on which the NEC holds,
then there are no compact future weakly trapped submanifolds
within $\doc$.
\end{Theorem}

\begin{Remark}
{\em Theorems \ref{notrapped0} and \ref{notrapped} may be
adapted to rule out the visibility from infinity of
submanifolds $S$ bounding compact acausal hypersurfaces $\hyp$ with
weakly  {\it outer} future trapped boundary. In fact $\hyp$ is
allowed to have non-weakly outer trapped components of the boundary
as long as those lie in a black hole region. Here the outer direction at $S$
is defined as pointing away from $\hyp$.}
\end{Remark}

\appendix
\section{Uniform boundaries in Lindblad-Rodnianski-Loizelet
metrics} \label{SALRL}

In this appendix we wish to point out that sufficiently small
data vacuum space-times constructed using the
Lindblad-Rodnianski method~\cite{LindbladRodnianski2}, as
generalised by Loizelet to higher
dimensions~\cite{Loizelet:these,Loizelet:AFT}
(compare~\cite{CCL}), contain past inwards trapped,
\emph{closed to the future} (in a sense which should be made
clear by what is said below),
timelike hypersurfaces. This is irrelevant as far as the
topological implications of our analysis are concerned, as in
this case the space-time manifold is $\R^{n+1}$ anyway, but it
illustrates the fact that such hypersurfaces can arise in
vacuum space-times which are not necessarily stationary. Note
that the resulting space-times are uniformly asymptotically
flat  to order zero,  but \emph{not} to order one in general,
as the retarded-time derivatives of a radiating metric will not
fall-off faster than $1/r$ when approaching future null
infinity.

In order to proceed, we recall some facts about the space-times
constructed in~\cite{LindbladRodnianski2,Loizelet:these}.
%
In Minkowski space-time $\R^{1,n}=(\R^{n+1},\eta)$ let
$$
 q= r-t
 \;,
$$
and let $H^{\mu\nu}:= g^{\mu\nu}-\eta^{\mu\nu}$, where
$g_{\mu\nu}$ is a small data vacuum  metric on $\R^{n+1}$ as
constructed in~\cite{LindbladRodnianski2,Loizelet:these}.
By~\cite[Corollary~9.3]{LindbladRodnianski2} for $n=3$,
and by \cite[Corollary 5.1]{Loizelet:these} for $n\ge 3$,%
\footnote{At the end of the bootstrap argument one concludes
that the inequalities there are satisfied by the solution.}
there exist constants $C$, $0<\delta<\delta'<1$
such that
%
\newcommand{\val}[1]{|#1|}%
\newcommand{\dbar}{\bar\partial}%
\begin{equation}\label{912}
\val{\partial  H}\leq
\begin{cases}
C\varepsilon(1+t+\val{q})^{ \frac{1-n}{2}
+\delta}(1+\val{q})^{-1-\delta'},\quad q\ge0\;,\\
C\varepsilon(1+t+\val{q})^{\frac{1-n}{2}+\delta}(1+\val{q})^{-1/2},\quad
q<0\;,
\end{cases}
\end{equation}
\begin{equation}\label{913}
\val{H}\leq
\begin{cases}
C\varepsilon(1+t+\val{q})^{\frac{1-n}{2}+\delta}(1+\val{q})^{-\delta'},\quad q\ge0\;,\\
C\varepsilon(1+t+\val{q})^{\frac{1-n}{2}+\delta}(1+\val{q})^{1/2},\quad
q<0\;,
\end{cases}
\end{equation}
\begin{equation}\label{914}
\val{\dbar{H}}\leq
\begin{cases}
C\varepsilon(1+t+\val{q})^{\frac{-1-n}{2}+\delta}(1+\val{q})^{-\delta'},\quad q\ge0\;,\\
C\varepsilon(1+t+\val{q})^{\frac{-1-n}{2}+\delta}(1+\val{q})^{1/2},\quad
q<0\;.
\end{cases}
\end{equation}
Here  $\epsilon$  and $\delta$ are  small constants determined
by the initial data, and $\delta$ can be chosen as small as
desired by choosing the data close enough to the Minkowskian
ones. Next, $\bar
\partial$ denotes partial coordinate derivatives $\partial_\mu$
to which a projection operator in directions tangent to the
outgoing coordinate cones $\{t-r=\const\}$ has been applied;
e.g., in spherical coordinates, $\bar
\partial \in \Span\{L:=\partial_t+\partial_r,\frac 1 r
\partial_\theta, \frac 1 {r\sin \theta} \partial_\varphi\}$.

Examining separately the cases $0\le t\le r/2$, $ r/2\le t \le
r$, and $r\le t$,  it is easily seen that there exists a
constant $C>0$
such that, for all $n\ge 3$,%
\footnote{The estimates are actually better in higher
dimensions, which is irrelevant for our purposes here.}
\bel{Hunifzero}
 |H|  \le \frac C {(1+r)^{1/2-\delta}}
\;,\quad
 |\partial H| \le \frac C  {(1+r)^{1-\delta}}
\;,
 \quad |\bar \partial H| \le \frac C {(1+r)^{3/2-\delta}}
 \;.
\ee
The first inequality implies that $(\mcM,\fourg)$ is uniformly
asymptotically flat to order zero. On the other hand,
$(\mcM,\fourg)$  is \emph{not} uniformly asymptotically flat to
order one. However, the third inequality shows that one can
choose $R_0$ large enough so that for all $R\ge R_0$ the
hypersurfaces $\{r=R\;, t\ge 0\}$ are \emph{inward past null
convex}, in the sense that the level sets of $t$ within
$\{r=R\}$ have negative definite past inwards null second
fundamental form (compare~\cite{Galloway:fitopology}). Indeed,
from the first inequality one can choose null normals to
$\{r=R\}$ of the form $\pm
\partial_0 \pm \partial_r +O(r^{1/2-\delta})$, with a uniform
error term. It now follows from the second and third estimate
that the null second fundamental forms differ from their
Minkowskian counterparts by terms which are uniformly
$O(r^{-3/2+2\delta})$ and $O(r^{-3/2+\delta})$. For
sufficiently small initial data one can choose $\delta<1/4$,
and the result follows.

A corresponding result holds for $t<0$ by invariance of the
Einstein equations under the map $t\mapsto -t$.

In particular the traces of the null second fundamental forms
have the right signs for the results in our work to apply.

\bigskip

\noindent {\sc Acknowledgements:} PTC and GG wish to thank David
Maxwell for useful discussions about the problem addressed
in Section~\ref{swftsi}.

\bibliographystyle{amsplain}
\bibliography
{
../../prace/references/newbiblio,%
../../prace/references/newbib,%
../../prace/references/reffile,%
../../prace/references/bibl,%
../../prace/references/Energy,%
../../prace/references/hip_bib,%
../../prace/references/netbiblio,../../prace/references/addon}

\def\cprime{$'$} \def\cprime{$'$} \def\cprime{$'$} \def\cprime{$'$}
\providecommand{\bysame}{\leavevmode\hbox to3em{\hrulefill}\thinspace}
\providecommand{\MR}{\relax\ifhmode\unskip\space\fi MR }
\providecommand{\MRhref}[2]{%
  \href{http://www.ams.org/mathscinet-getitem?mr=#1}{#2}
}
\providecommand{\href}[2]{#2}
\begin{thebibliography}{10}

\bibitem{BTZ}
M.~Ba{\~n}ados, C.~Teitelboim, and J.~Zanelli, \emph{{Black hole in
  three-dimensional space-time}}, Phys.\ Rev.\ Lett. \textbf{69} (1992),
  1849--1851, arXiv:hep-th/9204099.

\bibitem{BMG}
P.~Breitenlohner, D.~Maison, and G.~Gibbons, \emph{{$4$}-dimensional black
  holes from {K}aluza-{K}lein theories}, Commun.\ Math.\ Phys. \textbf{120}
  (1988), 295--333. \MR{MR973537 (89j:83018)}

\bibitem{CadeauWoolgar}
C.~Cadeau and E.~Woolgar, \emph{New five dimensional black holes classified by
  horizon geometry, and a {Bianchi VI} braneworld}, Class.\ Quantum Grav.
  (2001), 527--542, arXiv:gr-qc/0011029.

\bibitem{CarterlesHouches}
B.~Carter, \emph{Black hole equilibrium states}, Black Holes (C.\ de~Witt and
  B.\ de~Witt, eds.), Gordon \& Breach, New York, London, Paris, 1973,
  Proceedings of the Les Houches Summer School.

\bibitem{CCL}
Y.~Choquet-Bruhat, P.T. Chru\'{s}ciel, and J.~Loizelet, \emph{{Global solutions
  of the Einstein--Maxwell equations in higher dimension}}, Class.\ Quantum
  Grav. (2006), 7383--7394, arXiv:gr-qc/0608108.

\bibitem{ChHighDim}
P.T. Chru\'{s}ciel, \emph{On higher dimensional black holes with abelian
  isometry group}, Jour.\ Math.\ Phys. \textbf{50} (2008), 052501 (21 pp.),
  arXiv:0812.3424 [gr-qc].

\bibitem{ChCo}
P.T. Chru\'{s}ciel and J.~Costa, \emph{On uniqueness of stationary black
  holes}, Ast\'erisque (2008), 195--265, arXiv:0806.0016v2 [gr-qc].

\bibitem{ChDGH}
P.T. Chru\'{s}ciel, E.~Delay, G.~Galloway, and R.~Howard, \emph{Regularity of
  horizons and the area theorem}, Annales Henri Poincar\'e \textbf{2} (2001),
  109--178, arXiv:gr-qc/0001003. \MR{MR1823836 (2002e:83045)}

\bibitem{ChWald1}
P.T. Chru\'{s}ciel and R.M. Wald, \emph{Maximal hypersurfaces in stationary
  asymptotically flat space--times}, Commun.\ Math.\ Phys. \textbf{163} (1994),
  561--604, arXiv:gr--qc/9304009. \MR{MR1284797 (95f:53113)}

\bibitem{ChWald}
\bysame, \emph{On the topology of stationary black holes}, Class.\ Quantum
  Grav. \textbf{11} (1994), no.~12, L147--152, arXiv:gr--qc/9410004.
  \MR{MR1307013 (95j:83080)}

\bibitem{Coquereaux:1988ne}
R.~Coquereaux and A.~Jadczyk, \emph{{Riemannian geometry, fiber bundles,
  Kaluza-Klein} theories and all that}, World Sci.\ Lect.\ Notes Phys.,
  vol.~16, World Scientific Publishing Co., Singapore, 1988. \MR{MR940468
  (89e:53108)}

\bibitem{FriedmanSchleichWitt}
J.L. Friedman, K.~Schleich, and D.M. Witt, \emph{Topological censorship}, Phys.
  Rev. Lett. \textbf{71} (1993), 1486--1489, erratum {\textbf{75}} (1995) 1872.

\bibitem{galloway-topology}
G.J. Galloway, \emph{On the topology of the domain of outer communication},
  Class.\ Quantum Grav. \textbf{12} (1995), L99--L101.

\bibitem{Galloway:fitopology}
\bysame, \emph{A ``finite infinity'' version of the {FSW} topological
  censorship}, Class.\ Quantum Grav. \textbf{13} (1996), 1471--1478.
  \MR{MR1397128 (97h:83065)}

\bibitem{Galloway:splitting}
\bysame, \emph{Maximum principles for null hypersurfaces and null splitting
  theorems}, Ann. Henri Poincar\'e \textbf{1} (2000), 543--567. \MR{1 777 311}

\bibitem{GregCargese}
\bysame, \emph{Null geometry and the {E}instein equations}, The Einstein
  equations and the large scale behavior of gravitational fields, Birkh\"auser,
  Basel, 2004, pp.~379--400. \MR{MR2098922 (2006f:83015)}

\bibitem{GSWW}
G.J. Galloway, K.~Schleich, D.M. Witt, and E.~Woolgar, \emph{Topological
  censorship and higher genus black holes}, Phys. Rev. \textbf{D60} (1999),
  104039, arXiv:gr-qc/9902061.

\bibitem{GSWW2}
\bysame, \emph{The {AdS/CFT} correspondence conjecture and topological
  censorship}, Phys.\ Lett. \textbf{B505} (2001), 255--262,
  arXiv:hep-th/9912119.

\bibitem{HHM}
B.C. Haggman, G.W. Horndeski, and G.~Mess, \emph{Properties of a covering space
  defined by {H}awking}, Jour.\ Math. Phys. \textbf{21} (1980), 2412--2416.

\bibitem{Hatcher}
A.~Hatcher, \emph{Algebraic topology}, Cambridge University Press, Cambridge,
  2002. \MR{MR1867354 (2002k:55001)}

\bibitem{Jacobson:venkatarami}
T.\ Jacobson and S.\ Venkatarami, \emph{Topology of event horizons and
  topological censorship}, Class.\ Quantum Grav. \textbf{12} (1995),
  1055--1061.

\bibitem{Lee:TM}
J.M. Lee, \emph{Introduction to topological manifolds}, Graduate Texts in
  Mathematics, vol. 202, Springer-Verlag, New York, 2000. \MR{MR1759845
  (2001d:57001)}

\bibitem{Lee:SM}
\bysame, \emph{Introduction to smooth manifolds}, Graduate Texts in
  Mathematics, vol. 218, Springer-Verlag, New York, 2003. \MR{MR1930091
  (2003k:58001)}

\bibitem{LindbladRodnianski2}
H.~Lindblad and I.~Rodnianski, \emph{The global stability of the {Minkowski}
  space-time in harmonic gauge},  (2004), arXiv:math.ap/0411109.

\bibitem{Loizelet:these}
J.~Loizelet, \emph{Probl\`emes globaux en relativit\'e g\'en\'erale}, Ph.D.
  thesis, Universit\'e de Tours, 2008,
  \url{www.phys.univ-tours.fr/~piotr/papers/TheseTitreComplet.pdf}.

\bibitem{Loizelet:AFT}
\bysame, \emph{Solutions globales d'\'equations {Einstein Maxwell}}, Ann.\
  Fac.\ Sci.\ Toulouse (2008), in press.

\bibitem{Solis}
D.A. Solis, \emph{Global properties of asymptotically de {Sitter and Anti de
  Sitter} space-times}, Ph.D. thesis, University of Miami, 2006.

\end{thebibliography}
\end {document}